\documentstyle[prl,twocolumn,aps,psfig,epsf]{revtex}
\begin{document}                
\def\a{$\alpha$}
\def\be{\begin{equation}}
\def\ee{\end{equation}}
\def\ba{\begin{eqnarray}}
\def\ea{\end{eqnarray}}
\title{Quantization of a billiard model for interacting particles}
\author{Thomas Papenbrock\cite{INT} and Toma\v z Prosen\cite{FMP}}
\address{Centro Internacional de Ciencias A.C., 62131 Cuernavaca, Mexico}
\maketitle
\begin{abstract}
We consider a billiard model of a self-bound, interacting three-body system 
in two spatial dimensions. Numerical studies show that the classical dynamics 
is chaotic. The corresponding quantum system displays spectral fluctuations 
that exhibit small deviations from random matrix theory predictions. These
can be understood in terms of scarring caused by a 1-parameter family of 
orbits inside the collinear manifold.  
\end{abstract}
Billiards are interesting and useful models to study the quantum mechanics of 
classically chaotic systems \cite{Bohigas,BillCon}. In particular, the
study of the completely chaotic Sinai billiard \cite{Sinai} and 
Bunimovich stadium \cite{Stadium} showed that the quantum spectra and 
wave functions of classically chaotic system exhibit universal properties 
(e.g. spectral fluctuations) \cite{BoGiaSchmi} 
as well as deviations (e.g. scars of periodic orbits) \cite{Heller} 
when compared to random matrix theory (RMT) predictions of the 
Gaussian orthogonal ensemble (GOE) \cite{GMW}.

While the theory of wave function scarring has reached a mature state
in two dimensional systems \cite{Bogomolny,Berry,Agam,Kaplan} there
is a richer structure in more than two dimensions. In particular, invariant
manifolds in billiards \cite{ProsenI} and systems of identical particles
\cite{PapSel} may lead to an enhancement in the amplitude of wave functions
\cite{ProsenII,PSW} provided classical motion is not too unstable in their
vicinities. 

The purpose of this letter is twofold. First we want to quantize a billiard
model for three interacting particles and study a new type of wave function
scarring. Second, our numerical results strongly suggest that the system under 
consideration is chaotic and ergodic. This is interesting in view
of recent efforts to construct higher-dimensional chaotic billiards 
\cite{Primack95,BuCaGua96,BuRa97}. 

This letter is organized as follows. First we introduce a billiard model
of an interacting three-body system and study its classical dynamics.
Second we compute  highly excited eigenstates of the corresponding quantum 
system and compare the results with RMT predictions.

Recently, a self-bound many-body system realized as a billiard has been
studied in the framework of nuclear physics \cite{TP1}. We want to consider
the corresponding three-body system with the Hamiltonian
\be
\label{ham}
H=\sum_{i=1}^3{\vec{p}_i^2\over 2m} + \sum_{i<j}V(|\vec{r}_i-\vec{r}_j|),
\ee
where $\vec{r}_i$ is a two--dimensional position vector of the $i$-th particle
and $\vec{p}_i$ is its conjugate momentum. The two-body potential is
\ba
\label{int}
V(r)=\left\{
     \begin{array}{ll}
     0 & \mbox{for $r<a$}, \\
     \infty & \mbox{for $r\ge a$}.
     \end{array}
     \right.
\ea
The particles thus move freely within a convex billiard in six-dimensional
configuration space and undergo elastic reflections at the walls. Besides 
the energy $E$, the total momentum $\vec{P}$ and angular momentum $L$ 
are conserved quantities which leaves us with three degrees of freedom. 
In what follows we consider the case $\vec{P}=0, L=0$. 

To study the classical dynamics it is convenient to fix the velocity 
$\vec{v}_1^2+\vec{v}_2^2+\vec{v}_3^2=1$ and perform computations with the 
full Hamiltonian (\ref{ham}) without transforming to the subspace
$\vec{P}=0, L=0$.
We want to compute the Lyapunov exponents of several trajectories. 
To this purpose we draw initial conditions at random and  
compute the tangent map \cite{Sieber,BuCaGua96} while following their 
time evolution. To ensure good
statistics and good convergence we follow an ensemble of $7\times10^4$ 
trajectories for $10^5$ bounces off the boundary. All followed trajectories 
have positive Lyapunov exponents. The ensemble averaged value of the maximal
Lyapunov exponent and its RMS 
deviation are $\lambda a=0.3933 \pm0.0015$, while the second Lyapunov
exponent is also always positive. Thus, the system is chaotic 
for practical purposes. However, we have no general proof that no 
stable orbits exist. The reliability of the numerical
computation was checked by (i) comparing forward with backward evolution,
(ii) observing that energy, total momentum and angular momentum are conserved
to high accuracy during the evolution and (iii) using an alternative 
method \cite{Bennetin} to determine the Lyapunov exponent. 

\begin{figure}
  \begin{center}
    \leavevmode
    \parbox{0.3\textwidth}
           {\psfig{file=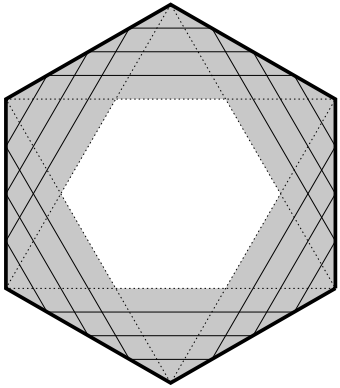,width=0.3\textwidth,angle=0}}
  \end{center}
\protect\caption{The motion inside the 
collinear manifold corresponds to the motion inside a hexagonal billiard. 
The parabolic-elliptic family of periodic orbits is shaded. Three of its
members and two limiting orbits are presented by full and dotted lines, 
respectively.}
\label{fig1}
\end{figure}

The considered billiard possesses two low-dimensional invariant manifolds
that correspond to symmetry planes. The first ``collinear'' manifold is 
defined by configurations where all three particles move on a line.
The dynamics inside this manifold is governed by the one-dimensional analogon
of Hamiltonian (\ref{ham}). After separation of the center-of-mass motion
one obtains a two-dimensional billiard with the shape of a regular hexagon.
This system is known to be {\em pseudo-integrable} 
\cite{Rich,Eck}. 
To study the motion in the vicinity of the collinear
manifold we compute the {\em full phase space} 
stability matrix for several periodic
orbits inside the collinear manifold which come in
1-dim families and can be systematically enumerated using the tiling 
property of the hexagon. All considered types of orbits {\em except 
two} are unstable in the {\em transverse} direction:
(i) The family of {\em bouncing ball} orbits (i.e. two particles bouncing, 
the third one
at rest in between) is marginally stable (parabolic) in full phase space. 
(ii) The family of orbits depicted in Fig.~\ref{fig1} 
is stable (elliptic) in two {\em transversal} directions and 
marginally stable (parabolic) in the other 10 directions of 
12-dim phase space. 
Though this behavior does not spoil the ergodicity of the billiard one may 
expect that it causes the quantum system to display deviations 
from RMT predictions. Note that this family of periodic orbits differs
from the bouncing ball orbits which have been extensively studied in two- and 
three-dimensional billiards \cite{Darmstadt,Primack95} since (i) it is 
restricted to a lower dimensional invariant manifold, and (ii) it is elliptic
(complex unimodular pair of eigenvalues) in one conjugate pair of directions. 

The second invariant manifold is defined by those
configurations where two particles are mirror images of each other while the
third particle is restricted to the motion on the (arbitrarily chosen)
symmetry line. Inside this manifold one finds mixed (i.e. partly regular and
partly chaotic) dynamics. However, the
motion is infinitely unstable in the transverse direction due to 
non-regularizable three-body collisions.

\begin{figure}
  \begin{center}
    \leavevmode
    \parbox{0.3\textwidth}
           {\psfig{file=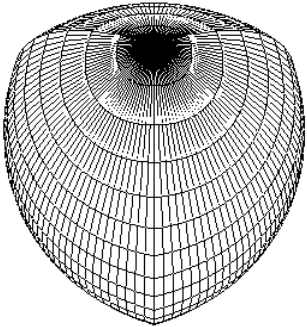,width=0.25\textwidth,angle=0}}
  \end{center}
\protect\caption{Billiard shape after separating off the center of mass
motion and rotational degree of freedom.}  
\label{fig2}
\end{figure}

The quantum mechanics is done using the coordinates
\ba
\vec{x} &=&\left(\vec{r_1}+\vec{r_2}+\vec{r_3}\right)/\sqrt{3},\nonumber\\
\rho\cos{\theta'\over2}\left(\begin{array}{c}
                       \cos{(\phi-\varphi'/2)}\\
                       \sin{(\phi-\varphi'/2)}\end{array}\right)
 &=& \left(\vec{r_1}-\vec{r_2}\right)/\sqrt{2},\nonumber\\
\rho\sin{\theta'\over2}\left(\begin{array}{c}
                       \cos{(\phi+\varphi'/2)}\\
                       \sin{(\phi+\varphi'/2)}\end{array}\right)
 &=& \left(\vec{r_1}+\vec{r_2}-2\vec{r_3}\right)/\sqrt{6}.
\label{trafo}
\ea
Here $\rho, \theta'$ and $\varphi'$ describe the intrinsic motion of the 
three-body system while $\vec{x}$ and $\phi$ are the center of mass and the
global orientation, respectively. 
In a second transformation we apply a rotation of 
$\pi/2$ around the abscissa corresponding to spherical coordinates 
$(\rho,\theta',\varphi')$, namely
$\tan{\varphi}=-\cot{\theta'}/\cos{\varphi'}$ and 
$\cos{\theta}=\sin{\theta'}\sin{\varphi'}$ and obtain for the
Laplacian in the subspace $\vec{P}=0,L=0$ 
\be
\label{laplace}
\Delta={\partial^2\over\partial\rho^2}+{3\over\rho}{\partial\over\partial\rho}
+{4\over\rho^2}\left({\partial^2\over\partial\theta^2}
+\cot{\theta}{\partial\over\partial\theta}+{1\over\sin^2{\theta}}
{\partial^2\over\partial\varphi^2}\right).
\ee
Products of Bessel functions and spherical harmonics 
\be
\psi_{k,l,l_z}(\rho,\theta,\varphi)=(k\rho)^{-1}\,J_{2l+1}(k\rho)\,
Y_l^{l_z}(\theta,\varphi)
\label{eq:sf}
\ee
are eigenfunctions of the Laplacian (\ref{laplace})
\be
\label{basis}
\Delta\psi_{k,l,l_z}(\rho,\theta,\varphi)
=-k^2\psi_{k,l,l_z}(\rho,\theta,\varphi),
\ee
with the usual relation between wavevector and energy 
$k=\hbar^{-1}(2mE)^{1/2}$.
Fig.~\ref{fig2} shows a picture of the billiard taking $(\rho,\theta,\varphi)$
as spherical coordinates. The billiard possesses a $D_{3h}$ symmetry.
In the fundamental domain 
$(\theta,\varphi)\in (0,\pi/2) \times (-\pi/6,\pi/6)$ the boundary is given
by
\be
\label{boundary}
\rho_B(\theta,\varphi)={a\over\sqrt{1+\sin{\theta}\sin{(\varphi+\pi/3)}}}.
\ee
The collinear manifold is the equatorial plane $\theta=\pi/2$. The second
invariant manifold is given by the vertical symmetry planes $\phi=\pm\pi/6$.
Note that in this representation, classical geodesics of the 
billiard between two successive collisions are not straight lines since 
the centrifugal potential is stronger than in Euclidean case.
In what follows we restrict ourselves to the fundamental domain and 
choose basis functions that fulfill Dirichlet boundary conditions.
These  are bosonic states.

We are interested in highly excited eigenstates. These may be 
accurately computed numerically by 
using the scaling method developed in ref. \cite{Saraceno} and applied to 
a three-dimensional billiard by one of the authors \cite{ProsenII}. 
This method works efficiently only when a suitable 
positive weight function is introduced in a boundary integral.
To this purpose we note that the radial part of (\ref{laplace}) looks 
like a 4-dim Laplacian. Extending the results of refs. 
\cite{Saraceno,ProsenII} to four dimensions yields the appropriate weight 
function, which has a remarkably simple form in our coordinates, namely
we minimize the following functional
$$
f[\Psi_k] = \int_0^1 d\cos\theta\int_{-\pi/6}^{\pi/6} d\varphi
\rho_B^4(\theta,\varphi)|\Psi_k(\rho_B(\theta,\varphi),\theta,\varphi)|^2
$$
where the wave-function is expressed in terms {\em scaling functions} 
(\ref{eq:sf}), $\Psi_k=\sum_{l} c_{l,l_z} \psi_{k,l,l_z}$. 
Due to our particular choice of boundary conditions we consider
only the terms for which $l+l_z$ is {\em odd} and $l_z=3m$, and truncate
at $l=l_{\rm max} = ka/2 + \Delta l \approx ka/2$.

We have computed three stretches of highly excited states. They consist of 
$7430$, $1813$ and $2362$ consecutive eigenstates with $120 < ka < 235$, 
$290 < ka < 300$ and $393 < ka < 400$, respectively. The last two stretches
comprise levels with sequential quantum numbers around $20000$ and $45000$, 
respectively.
The completeness of the series was checked by comparing the number of obtained
eigenstates with the leading order prediction from the Weyl-formula 
$\bar{d}(k)=c(24\pi^2)^{-1}(ak)^2,\; c\approx 0.51349$.

\begin{figure}
  \begin{center}
    \leavevmode
    \parbox{0.9\textwidth}
           {\psfig{file=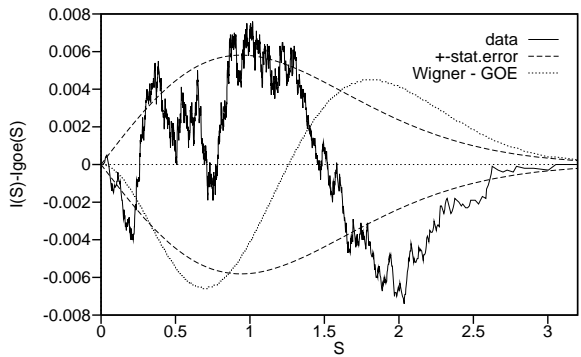,width=0.45\textwidth,angle=0}}
  \end{center}
\protect\caption{Integrated nearest neighbor spacing distribution 
$I(S)=\int_0^S ds P(S)$ with the GOE value subtracted,
$I(S)-I_{\rm goe}(S)$, for the set of $N=7430$ consecutive levels with
$120 < k < 235$ (full line). The dashed curve is the estimated statistical 
error $\sigma=\sqrt{I(S)(1-I(S))/N}$
and the dotted curve is the difference for the commonly used 
Wigner surmise $I_{\rm wig}(S)=1 - \exp(-\pi S^2/4)$.
}  
\label{fig3}
\end{figure}

Fig.~\ref{fig3} shows that the nearest neighbor spacing distribution 
agrees very well with RMT predictions already for the lower energy
spectral stretch $120 \le ka \le 235$. The other series show well agreement, 
too. 
As for the long-range spectral correlations, the number variance $\Sigma^2(L)$ 
deviates from RMT predictions for interval length of more than ten mean 
level spacings which we believe is due to the
parabolic-elliptic family of periodic orbits in the collinear manifold
(Fig.~\ref{fig1}). The deviation from RMT {\em decreases} with increasing $k$.
For the highest spectral stretch ($ka\approx 400$) the number variance 
increases linearly, 
$\Sigma^2(L) \approx \Sigma^2_{\rm goe}(L) + \varepsilon L$ up to
$L \le 250$, with $\varepsilon\approx 0.04$. This finding is consistent with 
the model of a statistically independent fraction $\varepsilon$ of 
strongly scarred states \cite{ProsenII}.
$\Sigma^2(L)$ reaches its maximum and begins to oscillate at the saturation 
length $L^*$ which scales as $L^*\propto k^2$ in agreement with the prediction 
of ref.~\cite{Berry400}.

\begin{figure}
  \begin{center}
    \leavevmode
    \parbox{0.9\textwidth}
           {\psfig{file=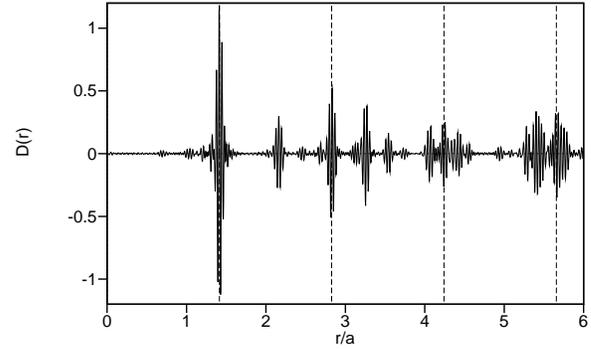,width=0.45\textwidth,angle=0}}
  \end{center}
\protect\caption{Length spectrum $D(r)$ for the long spectral sequence
with $120 < k < 235$. The dashed vertical lines denote the
integer multiples of the period of the parabolic-elliptic family.}  
\label{fig4}
\end{figure}

The length spectrum $D(r)$, i.e. the cosine transform of the oscillatory part 
of the spectral density $d_{\rm osc}(k)=\sum_n \delta(k-k_n)-\bar{d}(k)$, 
gives further information about long-range spectral fluctuations.
For finite stretches of consecutive levels in the interval $[k_1,k_2]$ one 
uses a Welsh window function $w(k;k_1,k_2) = (k_2-k)(k-k_1)/(6(k_2-k_1)^3)$ 
in the actual computation and obtains 
$D(r)=\int_{k_1}^{k_2} dk\, w(k;k_1,k_2) \cos{(kr)}\, d_{\rm osc}(k)$
(see e.g. \cite{Primack95,ProsenII}).
Fig.~\ref{fig4} shows that orbits of length $r=\sqrt{2}a$ and its integer
multiples cause dominant peaks in the length spectrum.

\begin{figure}[htbp]
\hbox{\hspace{-0.1in}
\vbox{
\vspace{6.54in}
\hbox{
\leavevmode
\epsfxsize=1.6in
\epsfbox{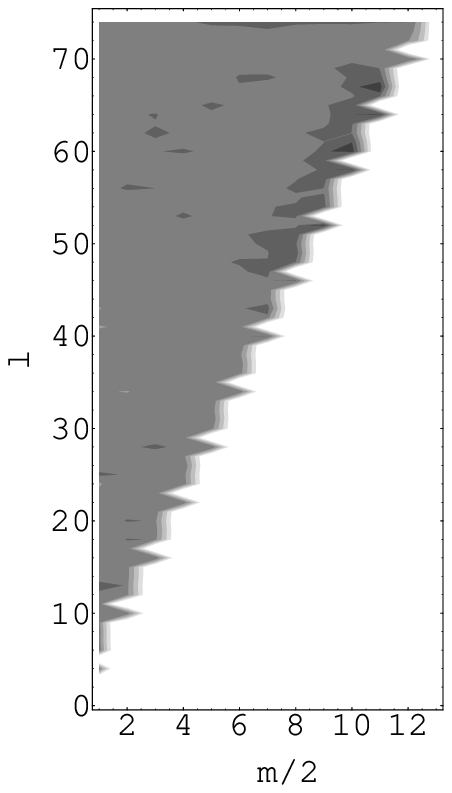}}
\vspace{-6.54in}}
\hspace{-0.2in}
\vbox{
\vspace{6.5in}
\hbox{
\leavevmode
\epsfxsize=1.6in
\epsfbox{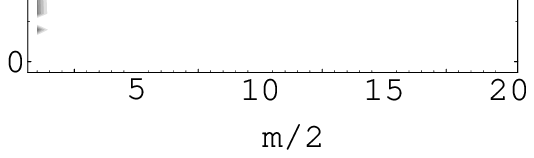}}
\vspace{-6.5in}
}}
\caption{${\rm IPR}(l,l_z=3m)$,
for 2052 states with $170 < ka < 200$ (left) and for 1813 states with 
$290 < ka < 300$ (right). 
Different levels of greyness are used for IPR values on consecutive
intervals of width $0.5$. The dominating light-grey corresponds to the
interval $[2.75,3.25]$ around the RMT value.}  
\label{fig5}
\end{figure}

To investigate the observed deviations from RMT predictions in more detail 
it is useful to compute the inverse participation ratio (IPR) of the 
wave functions in some basis \cite{Kaplan}. In the case of billiards 
the use of the angular momentum basis (\ref{eq:sf}) is particularly 
convenient and suitable 
since periodic orbits correspond to sets of isolated points within this 
representation. Let $c^{(n)}_{l,l_z}$ denote the expansion coefficients
of $n$th eigenstate $\Psi_{k_n}$. We compute the IPR over a set of $N$ 
consecutive eigenstates as 
${\rm IPR}(l,l_z)=N\sum_n|c^{(n)}_{l,l_z}|^4/(\sum_n|c^{(n)}_{l,l_z}|^2)^2$.
The predicted RMT value for ideally quantum ergodic states is 
${\rm IPR}_{\rm goe} = 3$. Fig.~\ref{fig5} shows the IPR for the two sets 
of eigenstates with $170 < k < 200$ and $290 < k < 300$, respectively.
The agreement with RMT predictions is rather good in both cases. 
This confirms that the billiard under consideration is dominantly chaotic 
and ergodic. However, the IPR is slightly enhanced in the region around 
$l = l_z\approx ka/3$. This is a robust phenomenon 
(present at all energy ranges), 
although the region of enhancement shrinks with increasing $k$. This finding
is compatible with the expectation of uniform quantum ergodicity in the 
semi-classical limit.
Note that the region $l\approx l_z$ corresponds to the vicinity 
of the collinear manifold. Note further that the orbits belonging to the 
parabolic-elliptic family depicted in Fig.~\ref{fig1}
have length $\sqrt{2}a$ and angular momenta in the region 
$l/ka=l_z/ka\in (1/2\sqrt{6},1/\sqrt{6})$. This is precisely the region
where the IPR exhibits its enhancement while the orbits' lengths coincide
with the prominent peaks of the length spectrum in Fig.~\ref{fig4}.

Thus, the deviations from RMT 
predictions observed for the spectrum and for the wave functions are 
associated with the family of parabolic-elliptic periodic orbits inside 
the collinear manifold. The special stability properties of 
this family lead  to scars in the wave functions of the quantum system. 
This is an exciting new type of scars of invariant manifolds and complements 
results previously found in a three-dimensional billiard \cite{ProsenII} 
and in interacting few-body systems \cite{PSW}. 
Note that the family of parabolic bouncing ball orbits inside the collinear 
manifold does not cause statistically detectable scarring. 
The orbits of this family correspond to points in angular momentum space 
with $l=l_z < ka/2\sqrt{6}$ and do not exhibit an enhancement in the IPR
since the classical motion is too unstable in their vicinity.

In summary we have investigated an interacting three-body system realized 
as a billiard. Numerical results show that the classical dynamics is 
dominantly chaotic and no deviation from ergodic behavior is found.
The spectral fluctuations of the quantum system agree well with random 
matrix theory predictions on energy scales of a few mean level spacings.  
However, wave function intensities and long ranged spectral fluctuations 
display deviations. These can be explained in terms of scars of a family of 
periodic orbits inside the collinear manifold. 

We thank L. Kaplan for stimulating discussions. 
The hospitality of the Centro Internacional de Ciencias, 
Cuernavaca, Mexico, is gratefully acknowledged.

\begin{references}  
\bibitem[a]{INT}
address: Institute for Nuclear Theory, Department of Physics, 
University of Washington, Seattle, WA 98195, USA.
e-mail: papenbro@phys.washington.edu
%
\bibitem[b]{FMP}
address: Physics Department, Faculty of Mathematics and Physics, 
University of Ljubljana, Jadranska 19, 1111 Ljubljana, Slovenia.
e-mail: prosen@fiz.uni-lj.si
%
\bibitem{Bohigas}
O. Bohigas and M.-J. Giannoni,
{\it Lecture Notes in Physics} Vol. 209 (Springer, Berlin 1984)
%
\bibitem{BillCon}
Proceedings of {\it Symposium on Classical and Quantum Billiards},
J. Stat. Phys. {\bf 83}, 1 (1996)
%
\bibitem{Sinai}
Y. A. Sinai,
Funct Anal. Appl. {\bf 2}, 61 and 245 (1968)
%
\bibitem{Stadium}
L. A. Bunimovich,
Funct Anal. Appl. {\bf 8}, 254 (1974)
%
\bibitem{BoGiaSchmi}
O. Bohigas, M.--J. Giannoni, and C. Schmit,
\prl {\bf 52} 1 (1984)   

\bibitem{Heller}
E. J. Heller, 
\prl {\bf 53} 1515 (1984)
%
\bibitem{GMW} 
T. Guhr, A. M\"uller-Groeling, and H. A. Weidenm\"uller,
Phys. Rep. {\bf 299}, 189 (1998) 
%
\bibitem{Bogomolny}
E. Bogomolny,
Physica {\bf D31}, 169 (1988)
%
\bibitem{Berry}
M. V. Berry,
Proc. R. Soc. Lond. {\bf A 423}, 219 (1989)
%
\bibitem{Agam}
O. Agam and S. Fishman,
\prl {\bf 73}, 806 (1994)
%
\bibitem{Kaplan}
L. Kaplan,
\prl {\bf 80}, 2582 (1998), Nonlinearity {\bf 12}, R1 (1999)

\bibitem{ProsenI}
T. Prosen,
Phys. Lett. {\bf A} 233, 323 (1997)
%
\bibitem{PapSel}
T. Papenbrock and T. H. Seligman,
Phys. Lett. {\bf A} 218, 229 (1996)
%
\bibitem{ProsenII}
T. Prosen,
\pl {\bf A} 233, 332 (1997)
%
\bibitem{PSW}
T. Papenbrock, T. H. Seligman, and H. A. Weidenm\"uller,
\prl {\bf 80}, 3057 (1998) 
%
\bibitem{Primack95}
H. Primack and U. Smilansky,
\prl {\bf 74}, 4831 (1995)
%
\bibitem{BuCaGua96}
L. Bunimovich, G. Casati, and I. Guarneri,
\prl {\bf 77}, 2914 (1996)
%
\bibitem{BuRa97}
L. A. Bunimovich and J. Rahacek, 
Com. Math. Phys. {\bf 189}, 729 (1997)
%
\bibitem{TP1}
T. Papenbrock, 
``Collective and chaotic motion in self-bound many-body systems'',
preprint DOE/ER/40561-50 (1999)
%
\bibitem{Sieber} 
M. Sieber and F. Steiner,
Physica {\bf D44}, 248 (1990)
%
\bibitem{Bennetin} 
G. Bennetin, L. Galgani, and J.M. Strelcyn,
\pra {\bf 14}, 2338 (1976)
%
\bibitem{Rich}
P. J. Richens and M. V. Berry,
Physica D {\bf 2} 495 (1981)
%
\bibitem{Eck}
B. Eckhardt, J. Ford, and F. Vivaldi,
Physica D {\bf 13}, 339 (1984)
%
\bibitem{Darmstadt}
H.-D. Gr\"af {\it et al.},
\prl {\bf 69}, 1296 (1992)
%
\bibitem{Saraceno}
E. Vergini and M. Saraceno,
\pre {\bf 52}, 2204 (1995)
%
\bibitem{Berry400}
M.V. Berry, Proc. Roy. Soc. Lond. A{\bf 400}, 229 (1985)
%
\end{references}
\end{document}